\documentclass[aps,prl,superscriptaddress,amsmath,amssymb,twocolumn,a4paper,floatfix]{revtex4-1}
\usepackage{graphicx}
\usepackage{float}
\usepackage{hyperref}
\usepackage{bm}
\usepackage{tabularx}
\usepackage{xcolor}
\usepackage{bbm}
\usepackage{accents}

\newcolumntype{Z}{>{\centering\let\newline\\\arraybackslash\hspace{0pt}}X}

\newcommand{\tr}{\text{tr}}

\begin{document}
\title{Vertex-based Diagrammatic Treatment of Light-Matter-Coupled Systems}
\author{Aaram J. Kim}
\affiliation{Department of Physics, University of Fribourg, 1700 Fribourg Switzerland}
\author{Katharina Lenk}
\affiliation{Department of Physics, University of Erlangen-N\"urnberg, 91058 Erlangen, Germany}
\author{Jiajun Li}
\affiliation{Department of Physics, University of Fribourg, 1700 Fribourg Switzerland}
\affiliation{Paul Scherrer Institute, Condensed Matter Theory, PSI Villigen, Switzerland}
\author{Philipp Werner}
\affiliation{Department of Physics, University of Fribourg, 1700 Fribourg Switzerland}
\author{Martin Eckstein}
\affiliation{Department of Physics, University of Erlangen-N\"urnberg, 91058 Erlangen, Germany}

\begin{abstract}
	We propose a 
diagrammatic Monte Carlo approach for 
general spin-boson models, which can be regarded as 
a generalization of
the strong-coupling expansion for fermionic impurity models. 
The algorithm is based on a self-consistently computed three-point vertex and a stochastically sampled four-point vertex, and achieves convergence to the numerically exact result in a wide parameter regime. The 
performance 
of the algorithm is demonstrated with applications to a spin-boson model representing an emitter in a waveguide. As a function of the coupling strength, the spin exhibits a delocalization-localization crossover at low temperatures, signaling a qualitative change in the real-time relaxation. 
In certain parameter regimes, the response functions of the emitter coupled to the electromagnetic continuum can be described by an effective  
Rabi 
model with appropriately defined parameters.
We also discuss the spatial distribution of the 
photon density around the emitter.
\end{abstract}

\maketitle

{\it Introduction.} The optical control of matter has emerged as a promising pathway for tuning material properties. 
The established paradigm involves disturbing materials with strong lasers  \cite{Basov2017}, leading e.g., to exotic nonthermal phases  \cite{Fausti2011a, McIver2020, Yamakawa2021}.  
Recently, the enhancement of the coupling between matter and vacuum fluctuations of the electromagnetic field in cavities has been identified as an alternative route for simultaneously engineering matter and light. When photon modes are confined in a near-field cavity or a waveguide, the hybridization between the material and photons can become significantly enhanced, giving rise to novel light-matter phases which embody unusual electronic and optical properties \cite{kockum2019, forn-diaz2019}. Possible applications include controlling the rates of chemical reactions through strong collective light-matter coupling \cite{ebbesen2016}. In the context of solid-state physics, experiments 
have revealed that strong quantum light-matter coupling can enhance ferromagnetism \cite{thomas2021}, change the robustness of topological phases \cite{appugliese2021}, and possibly affect the critical temperature of unconventional superconductors \cite{thomas2019}. 

A systematic theoretical analysis of these  scenarios is challenging. In particular, matter can generically interact with a continuum of photon modes \cite{sheremet2021, forn-diaz2017}, such as dispersive waveguide modes \cite{rokaj2020}, which are selectively enhanced by the optical confinement. 
Ultrastrong coupling between single emitters and an electromagnetic continuum has been experimentally realized in circuit quantum electrodynamics using superconducting qubits~\cite{Blais2004}, and similar physics can be studied by coupling a dipole to quantized surface acoustic waves \cite{manenti2017}. Moreover, effective strong coupling may be realized by exploiting an intermediate layer of excitations collectively coupled to both the dipole and photons \cite{schuetz2020,sidler2020}. 
Previous studies have attacked the problem with  polaron transformations \cite{diaz-camacho2016, shi2018} and matrix-product state simulations  \cite{sanchez-burillo2014} for a one-dimensional transmission line cavity. For general cavity setups, perturbative expansions around the high cavity frequency and the infinitely strong coupling limit have  been used to go beyond weak-coupling theory \cite{schlawin2018,li2021,ashida2021}. However, a theoretical or numerical tool capable of obtaining an unbiased description of matter which is strongly coupled to a photon continuum is still lacking. 

In this paper, we introduce a diagrammatic approach based on the self-consistent computation of a triangular vertex and the numerical evaluation of a four-point vertex, which is conceptually related to the established pseudo-particle methods \cite{Keiter1971,Pruschke1989,Haule2001} for fermionic quantum impurity models, 
and 
allows to obtain numerically exact results. To demonstrate the usefulness of the approach, 
we apply 
it  
to a strongly coupled spin-boson model, representing an emitter in a waveguide. 
We compute the delocalization-to-localization crossover as a function of light-matter coupling strength and 
analyze how the behavior of the dipole can be reproduced by a single-mode spin-boson model with appropriately defined effective parameters.

\begin{figure}[t]
	\centering
	\includegraphics[width=0.42\textwidth]{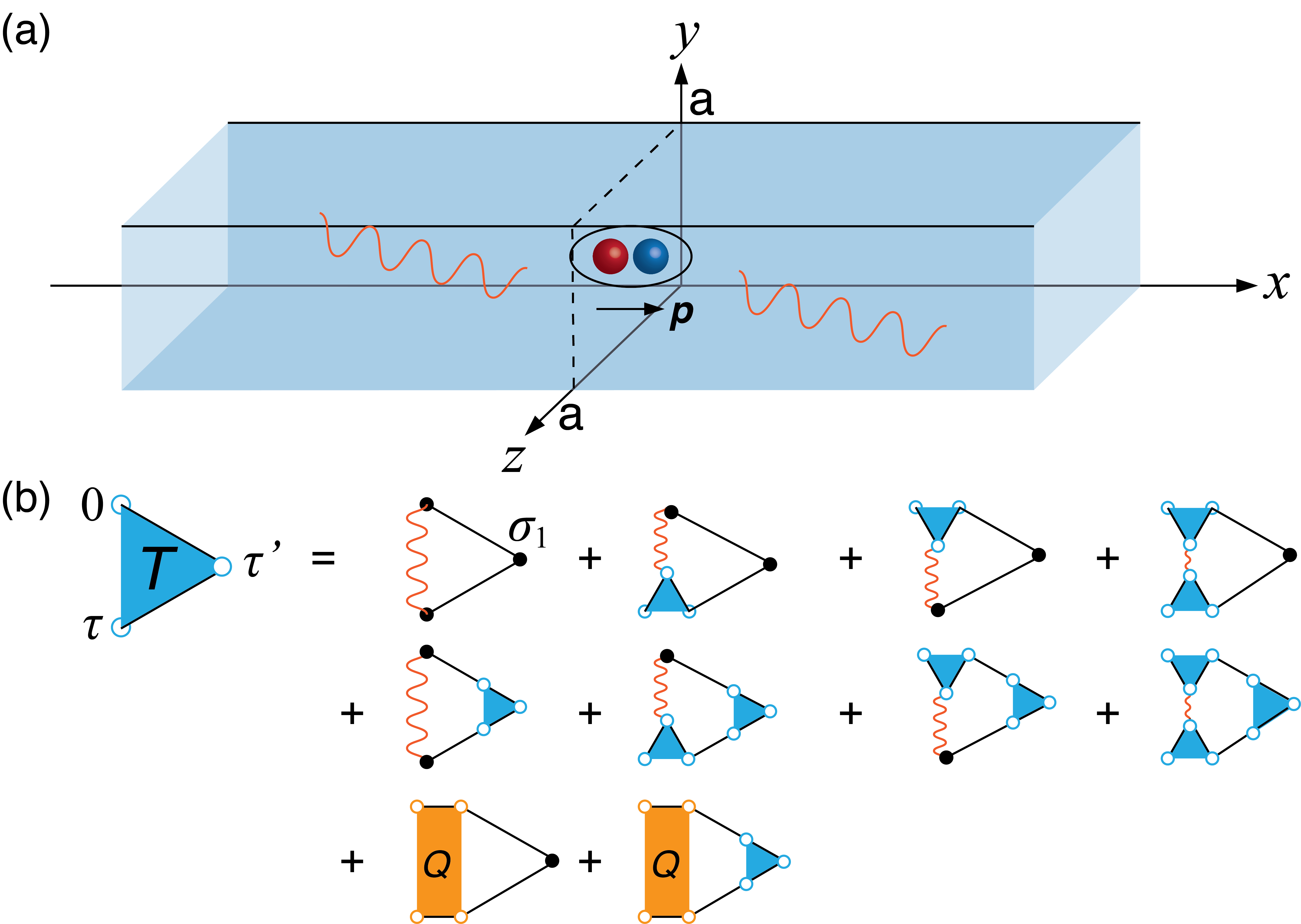}
	\caption{
		(a) Schematic picture of the rectangular waveguide with width $a$ in both the $y$ and $z$ direction. 
		The electric dipole with moment $\mathbf{p}=p\hat{\sigma}_1 {{\bf e}}_x$ along the $x$ direction is located at $(0,a/2,a/2)$.
		(b) Self-consistency equation for the triangular vertex $\hat{T}(\tau,\tau')$.
		The (red) wavy lines present the effective spin interaction $V(\tau-\tau')$ and (black) solid lines show the local propagator $\hat{\mathcal{G}}$.
		The (black) solid dot denotes the Pauli matrix $\hat{\sigma}_1$ and the open dots emphasize the encapsulated $\hat{\sigma}_1$ in the triangular (blue) and four-point (orange) vertex.
	}
	\label{fig:waveguide}
\end{figure}

{\it Model and Method.}
We consider an infinitely extended rectangular waveguide whose height and width are set to $a$. 
A two-level atom is located at  position $\left(a/2,a/2,0\right)$ and interacts with the confined photon modes through a dipolar coupling, see Fig.~\ref{fig:waveguide}(a). 
 In this geometry, the photon wavenumbers along the $y$ and $z$ direction ($k_y\equiv l\frac{\pi}{a},k_z\equiv m \frac{\pi}{a}$) are multiples of $\pi/a$ and will be denoted by $(k_l,k_m)$, while the $k_x$ component can take continuous values up to some cut-off frequency $\omega_c$. (We fix $\omega_c=30$ eV in the calculations.) 
The waveguide geometry generates a gap $\Omega_{11}=\sqrt{2}c\pi/a$ in the photon spectrum, with $c$ the speed of light. We use the notation $\Omega_{lm}^2=(k_l^2+k_m^2)c^2$, so that $\Omega_{11}$ denotes the photon energy for the lowest transverse wavenumber ($l=m=1$) and a vanishing $k_x$ component. In the main text, we only consider the the continuum near $\Omega_{11}$. 

In the dipolar gauge~\cite{Jiajun2020}, the Hamiltonian of the atom in the waveguide can be expressed as a spin-boson model,
\begin{equation}
	\hat{\mathcal{H}} = \frac{\Delta}{2}\hat{\sigma}_3 + \frac{1}{\sqrt{L}}\sum^{}_{k\gamma}\left(g_{k\gamma} \hat{\sigma}_1 {a}^{}_{k\gamma}+h.c.\right) + \sum^{}_{k\gamma}\omega_k {a}^{\dagger}_{k\gamma}{a}^{}_{k\gamma}~,
	\label{eqn:H}
\end{equation}
where $\hat{\sigma}_i$ ($i=1,2,3$) denotes the spin-$1/2$ Pauli operator in the basis of the two matter states, and $a^\dagger_{k\gamma}$ ($a_{k\gamma}$) is the photon creation (annihilation) operator with combined momentum index $k=(k_x,k_y,k_z)$ and (transverse) polarization mode index $\gamma=1,2$. $L$ is the normalization length of the waveguide along $x$ direction. 
The corresponding bare photon energy is $\omega_k=c\sqrt{k_x^2+k_y^2+k_z^2}$. $\Delta$ parametrizes the level splitting of the atomic states and we fix $\Delta=1.0$ eV.  The light-matter coupling  is given by $\sum_\gamma |g_{k\gamma}|^2=\frac{p^2\Omega_{11}^2}{\pi\epsilon a^2 \omega_k}$, where $p$ is the dipole matrix element for transitions between the two atomic states, and $\epsilon$ the vacuum permittivity.  By integrating out the photon degrees of freedom (see SM), one obtains an imaginary-time action with a retarded spin-spin interaction for inverse temperature $\beta$, 
\begin{equation}
	\mathcal{S} = \mathcal{S}_0 - \frac{1}{2}\int_{0}^{\beta}d\tau\int_{0}^{\beta}d\tau'~\hat{\sigma}_1(\tau)V(\tau-\tau')\hat{\sigma}_1(\tau')~,
	\label{eqn:action}
\end{equation}
where $\mathcal{S}_0$ denotes the local spin action and $V(\tau)=\sum^{}_{k\gamma}|g_{k\gamma}|^2\mathcal{D}^{0}_{k}(\tau)$,  with the bare photon propagator for momentum $k$ given by $\mathcal{D}^{0}_{k}(\tau)=e^{-\omega_k\tau}\theta(-\tau)n_B(\omega_k)+e^{-\omega_k\tau}\theta(\tau)[1+n_B(\omega_k)]$; $n_B$ denotes the Bose distribution function and $\theta$ the Heaviside function.

The solution of the model is formulated in terms of  the resolvent operator (or pseudo-particle propagator) $\hat {\mathcal{G}}(\tau) = \text{tr}_{\mathrm{ph}}[ \mathcal{T}_\tau \exp(-\int_0^\tau d\tau' \hat{H}(\tau'))]$ in imaginary time ($0\le\tau\le\beta$), which is a $2\times 2$  matrix in matter space after the partial trace over the photon degrees of freedom. Starting from the noninteracting propagator $\hat{\mathcal{G}}_0(\tau)=e^{-\frac{\Delta}{2}\hat{\sigma}_3\tau}$, a systematic diagrammatic perturbation in the retarded spin-spin interaction $V(\tau)$ can be written down in terms of a self-energy correction $\hat \Sigma(\tau)$ and the  time-ordered Dyson equation $\hat{\mathcal{G}}(\tau) = \hat{\mathcal{G}}_0(\tau) + \int_{0}^{\tau}d\tau_2\int_{0}^{\tau_2}d\tau_1~\hat{\mathcal{G}}_0(\tau-\tau_2)\hat \Sigma(\tau_2-\tau_1) \hat{\mathcal{G}}(\tau_1)$. The exact self-energy is given by the  sum of the leading term $\hat\Sigma_{\mathrm{NCA}}(\tau) = V(\tau) \hat \sigma_1 \hat{\mathcal{G}} (\tau)\hat\sigma_1$, called non-crossing approximation (NCA), and the vertex correction $\hat \Sigma(\tau)=\int_{0}^{\tau}d\tau_1\int_{\tau_1}^\tau d\tau_2 \hat {\mathcal{G}}(\tau-\tau_2)V(\tau-\tau_1)\hat T(\tau_2,\tau_1)$. Here the three-point vertex $\hat T(\tau_2,\tau_1)$ sums up all diagrams with interaction lines dressing the operator $\hat \sigma_1$; it can be represented in terms of a self-consistent set of exact diagrammatic equations, as illustrated in Fig.~\ref{fig:waveguide}(b). 
In the figure, the 
(orange) square is the four point vertex $Q(\tau;\tau_2,\tau_1)$, 
which is one-particle irreducible in the interaction line and two-particle irreducible in the local time-evolution operator. The triangular vertex equation must be solved self-consistently with the Dyson equation for $\hat \Sigma$. To compute the $Q$ vertex, we developed a diagrammatic Monte Carlo (diagMC) scheme \cite{Prokofev1998,VanHoucke2010,Kozik:2010fla}, which stochastically samples all possible Feynman diagrams of the $Q$ vertex. Upon convergence with diagram order, the addition of the $Q$ vertex  
in the self-consistency equation for $\hat T$ guarantees a numerically exact solution. Finally, relevant observables are evaluated in terms of $\hat {\mathcal{G}}$ and $\hat T$; in particular, the exact spin-correlation function $\chi_\text{sp} = \langle \hat{\sigma}_1(\tau)\hat{\sigma}_1(0)\rangle$ is given by 
\begin{align}
	&\chi_\text{sp} (\tau)= \frac{1}{Z}\tr\left[ \hat{\mathcal{G}}(\beta-\tau)\hat{\sigma}_1\hat{\mathcal{G}}(\tau)\hat{\sigma}_1 \right]+ \frac{1}{Z}\int_{0}^{\tau}\!d\tau_1\int_{\tau}^{\beta}\!d\tau_2
	\nonumber\\
	&\,\,\times\,\,\,\tr\left[\hat{\mathcal{G}}(\beta-\tau_2)\hat T(\tau_2-\tau_1,\tau-\tau_1)\hat{\mathcal{G}}(\tau_1)\hat{\sigma}_1\right],
	\label{eqn:chisp}
\end{align}
with $Z=\text{tr}[\hat{\mathcal{G}}(\beta)]$. 
We will also compare the exact solution to simpler schemes which do not involve the four-point vertex, in particular the NCA approximation $\hat{\Sigma}\approx\hat{\Sigma}_{\mathrm{NCA}}$, the one-crossing approximation (OCA) and two-crossing approximation (TCA) 
which keep self-energy diagrams with one and two crossings of the interaction lines, respectively (OCA approximates $\hat T$ by the first diagram in Fig.~\ref{fig:waveguide}), and the triangular vertex approximation (TVA), corresponding to the full self-consistent solution at $Q=0$. 

\begin{figure}[t]
	\centering 
	\includegraphics[width=0.45\textwidth]{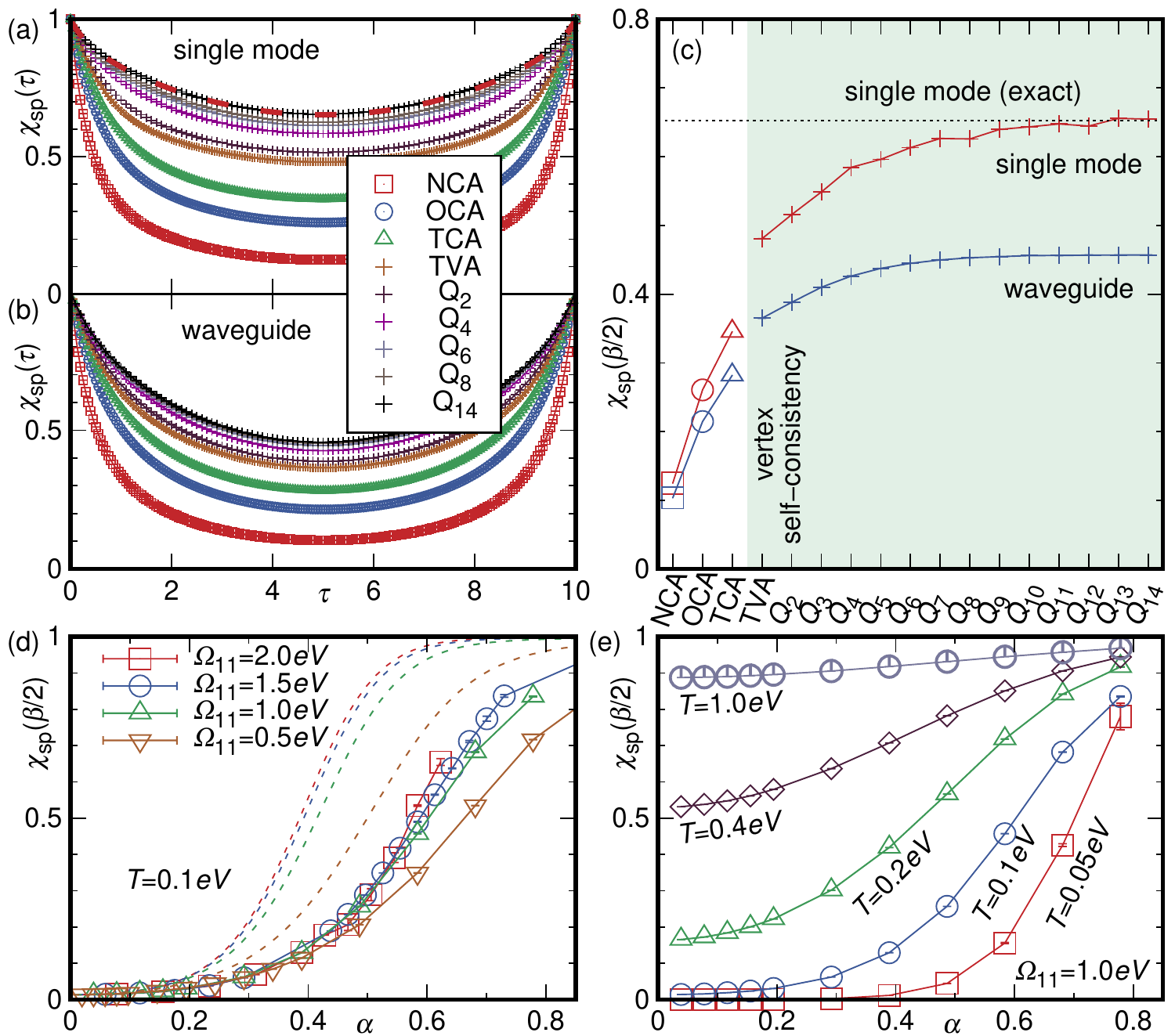}
	\caption{ 
Spin correlation function $\chi_{\mathrm{sp}}(\tau)$ for the approximate schemes (NCA, OCA, TCA, TVA) and for the Monte Carlo simulation including the $Q$-vertex diagrams to order $n$ (labelled $Q_n$); (a) Single mode benchmark with $T=0.1$ eV, $g=1.4$ eV, and $\omega=1.5$ eV; (b) Waveguide setup with $T=0.1$ eV, $p=3$ e$\mu$m, and $\Omega_{11}=1.0$ eV. The (red) dashed line in (a) presents the exact diagonalization results for the benchmark. (c) Systematic convergence of $\chi_{\mathrm{sp}}(\beta/2)$ with diagram order. (d,e) Converged spin correlation function for the waveguide model at $\tau=\beta/2$ as a function of the coupling strength $\alpha$, for various temperatures at fixed photon gap $\Omega_{11}=1.0$ eV (e), for various photonic gap sizes at a fixed temperature $T=0.1$ eV (d). Dashed lines in (d) show the results for the effective single-mode cavity model (see text). 
	}
	\label{fig:convergence}
\end{figure}

As a 
first 
test of the solver,  we consider the  model with only a single photon mode, with parameters $T=0.1$ eV, $\omega=1.5$ eV and $g=1.4$ eV, for which the spin-correlation function $\chi_\text{sp} = \langle \hat{\sigma}_1(\tau)\hat{\sigma}_1(0)\rangle$ can be calculated by exact diagonalization. In Fig.~\ref{fig:convergence}(a), one can see that the exact result (red dashed line) is recovered for sufficiently high ($\gtrsim 10$) diagram order in $Q$. Figure~\ref{fig:convergence}(c) shows that the applied vertex self-consistency (red crosses) considerably improves the results compared to schemes without it, like  
NCA, OCA, and TCA. 

{\it Results.} We now turn to the waveguide setup with parameters $T=0.1$ eV, $p=3$ $e\mu$m, and $\Omega_{11}=1.0$ eV.  
This parameter set represents the most challenging regime, where all energy scales are comparable. 
(For specific applications such as superconducting qubits~\cite{Blais2004}, one would simply have to rescale the energy unit.) 
Convergence to the exact result can be achieved by sampling the $Q$ vertex up to order $14$. The blue crosses in Fig.~\ref{fig:convergence}(c) illustrate the systematic convergence of $\chi_\text{sp}$  as a function of diagram order of the $Q$ vertex.  As in the single-mode case, the vertex self-consistency improves the accuracy, and the corrections from the $Q$ vertex are essential for reliable results in this strong-coupling regime.

At low but nonzero temperatures, the atom in the waveguide exhibits a crossover from a fluctuating state to a polarized state 
with increasing coupling strength, which we parametrize with the 
dimensionless $\alpha^2=\frac{p^2\Omega_{11}^2}{2\varepsilon\pi^3c^3}$ \footnote{
	Since $\Omega_{11}^2\propto 1/a^2$, we have $\tilde{g}^2_{k}/\omega_k=\sum_\gamma \tilde{g}^2_{k\gamma}/\omega_k \propto \Omega_{11}^2\Omega_{k}^2/\omega^2_k=\Omega_{11}^2/(1+c^2k_x^2/\Omega_{k}^2)$. 
	Setting $\omega_k=\Omega_k$ ($k_x=0$) leads to an effective coupling $\propto\Omega_{11}^2$. 
	It is worth noting that the sum $\sum_k \tilde{g}_k^2/\omega_{k}$ is the high-frequency limit of the cavity-photon contribution to the system energy, which generally grows slightly faster with $\Omega_{11}$ than in the single-mode approximation \cite{li2021}.
}. 
Figure~\ref{fig:convergence}(d,e) shows the converged $\chi_{\mathrm{sp}}(\beta/2)$, a measure for the ``localization" of the dipole, as a function of $\alpha$. Figure~\ref{fig:convergence}(e) plots $\chi(\beta/2)(\alpha)$ for different temperatures and a fixed photonic gap. As we enhance the quantum coherence by decreasing the temperature, 
$\chi_{\mathrm{sp}}(\beta/2)$  is considerably suppressed in the weak-coupling regime ($\alpha\lesssim 0.4$). In the 
strong-coupling regime $\chi_{\mathrm{sp}}$ shows a slow decay at long times, indicating a localized spin in the $x$ direction. The crossover defined by the inflection point of $\chi_{\mathrm{sp}}(\beta/2)$ gradually shifts to stronger couplings $\alpha$ and the crossover becomes sharper as we decrease the temperature.  Since the photon spectrum is gapped, and $V(\tau)$ decays faster than $\tau^{-2}$, we do not expect a localization transition at zero temperature \cite{Froehlich1982}.

For a fixed nonzero temperature, the $\chi_{\mathrm{sp}}(\beta/2)$ data for various photonic gap sizes collapse onto a single curve in the weak coupling regime (Fig.~\ref{fig:convergence}(e)), when plotted as a function of the dimensionless parameter $\alpha$. In the crossover regime, the $\chi_{\mathrm{sp}}$ curves for different gap sizes start to disperse; the larger the gap size, the lower the crossover point in terms of $\alpha$. The data, however, indicate a nontrivial crossover to a localized state in the limit $\Omega_{11}\rightarrow \infty$, where the photonic gap is much larger than the level splitting. This is because of the increasing light-matter coupling strength with increasing photonic gap (decreasing width $a$ of the waveguide).  

\begin{figure}[tbp]
	\centering
	\includegraphics[width=0.45\textwidth=270]{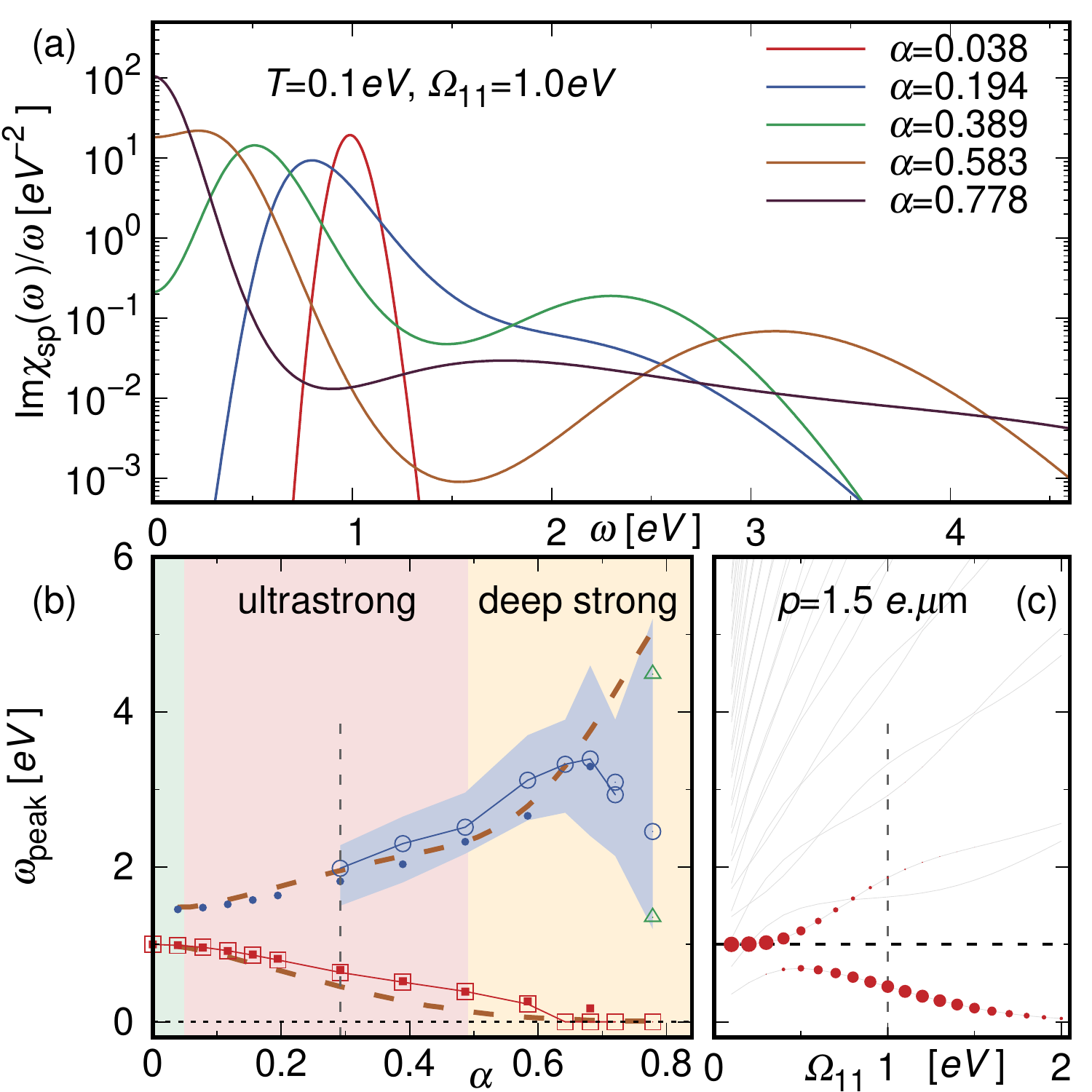}
		\caption{
		(a) Spin relaxation function $\mathrm{Im}\chi_{\mathrm{sp}}(\omega)/\omega$ for various coupling strengths $\alpha$, analytically continued by MaxEnt (fixed $T=0.1$ eV and $\Omega_{11}=1.0$ eV). (b) Peak locations of the lower and upper polariton modes obtained by MaxEnt (open symbols) and Pad\'e (solid symbols), with the full width at half-maximum shown by the blue shading. Brown dashed lines show the relevant excitation energies in the effective Rabi model (see text); for the given parameters, the couplings are $g_{\rm eff}=2.86\Omega_{11}\alpha$ and $\omega_{\rm eff}=1.4\Omega_{11} $. (c) Spectrum of the effective Rabi model, for $p=1.5$ e$\mu$m as a function of $\Omega_{11}$. For $\Omega_{11}=1$, the corresponding $\alpha=0.292$, see vertical line in panel (b). The symbol size is proportional to the contribution of the excitation to $\chi_{\mathrm{sp}}(\omega)$. 
}
	\label{fig:spectrum}
\end{figure}

Figure~\ref{fig:spectrum}(a) presents the spin relaxation function $\mathrm{Im}\chi_{\mathrm{sp}}(\omega)/\omega$, which we obtained by analytically continuing $\chi_\text{sp}(\tau)$ by the maximum entropy (MaxEnt) method \cite{Bryan1990,Gubernatis1996}. Without the waveguide, the result would be a delta-function at $\omega=1$ eV (broadened by MaxEnt). The spin-photon coupling splits the spin excitation into two separate polariton modes. The lower polariton mode shifts to $\omega\rightarrow 0$ as we increase $\alpha$, while the upper polariton mode 
moves to higher energy. 
This broad upper polariton mode is strongly enhanced with decreasing photonic gap, see SM. 
The inset of Fig.~\ref{fig:spectrum}(a) presents the peak positions of the lower and upper polariton modes  estimated by MaxEnt (open symbols) and Pad\'e \cite{Vidberg1977} (full symbols) analytical continuation.  The width of the high-energy satellite increases with increasing $\alpha$, and for  $\alpha\gtrsim 0.6$ it  becomes difficult to pinpoint the peak location within the numerical accuracy of analytic continuation. In the same strong coupling regime, the two low energy peaks at positive and negative energy start to merge, 
resulting in a single peak at $\omega=0$. This 
signals a qualitative change in the relaxation dynamics of the spin in real time, 
i.e., $\langle \hat{\sigma}_1(t)\rangle$, for a polarized initial condition $\langle \hat{\sigma}_1\rangle=1$ at $t=0$. In the  
case of a two-peak spectrum, the spin shows an underdamped oscillation, while it exhibits an overdamped relaxation in the single-peak case.

The splitting of the excitation spectrum into upper and lower polaritons looks similar to a conventional Rabi model, where a two-level emitter is coupled to a single cavity mode, even though the situation in the present case is very different, and for the parameters of Fig.~\ref{fig:spectrum}, the bare excitation energy $\Delta$ is located right at the edge of a continuum. Upon increasing the light-matter coupling, the lower polariton is pushed into the photonic gap (but still remains damped), while the upper polariton overlaps with the photon continuum. Nevertheless, the retarded interaction $V(\tau)$ suggests a way  to construct an effective Rabi model with photon energy $\omega_{\rm eff}$ and light-matter coupling  $g_{\rm eff}$, which can provide insights into the polariton splitting.  We determine the two effective parameters by identifying the retarded interaction $V(\tau)$ of the waveguide model with $V_{\mathrm{eff}}(\tau)=g_{\text{eff}}^2\cosh\left((\tau-\beta/2)\omega_{\text{eff}} \right)/\sinh\left(\beta\omega_{\text{eff}}/2\right)$ of the effective model at $\tau=0$ and $\beta/2$. The detailed dependence of $\omega_{\rm eff}$ and $g_{\rm eff}$ on $\alpha$ and $\Omega_{11}$ is analyzed in the SM. (In particular, in the low temperature limit $\beta \Omega \gg 1$, we have $\omega_{\rm eff}\approx \Omega_{11}$, and $g_{\rm eff}^2\approx 2\alpha^2\Omega_{11}^2\log(2\omega_c/\Omega_{11})$, with a logarithmic dependence on the high-energy cutoff.) In Fig.~\ref{fig:spectrum}(c), we show the excited states of the Rabi model with $n_\text{ph}\le 1$ photons, and their contribution to  $\chi_{\text{sp}}(\omega)$ as a function of $\Omega_{11}$ at fixed $p$. One can see that the lower polariton mode becomes the dominant spin excitation in  the narrow waveguide with $\Omega_{11}>\Delta$. It is renormalized towards zero for large $\Omega_{11}$, because of the increase of $g_{\rm eff}$ with $\Omega_{11}$. Finally, the dashed lines in Fig.~\ref{fig:spectrum}(b) show the two leading excitation energies in the resulting Rabi model, which fit the exact behavior remarkably well. This shows that the effective Rabi model provides a meaningful estimate of the coupling to a continuum, at least in the regime where the bare mode is not overlapping with the continuum. For a smaller value of $\Omega_{11}$, see SM.

\begin{figure}[t]
	\centering 
	\includegraphics[width=0.45\textwidth]{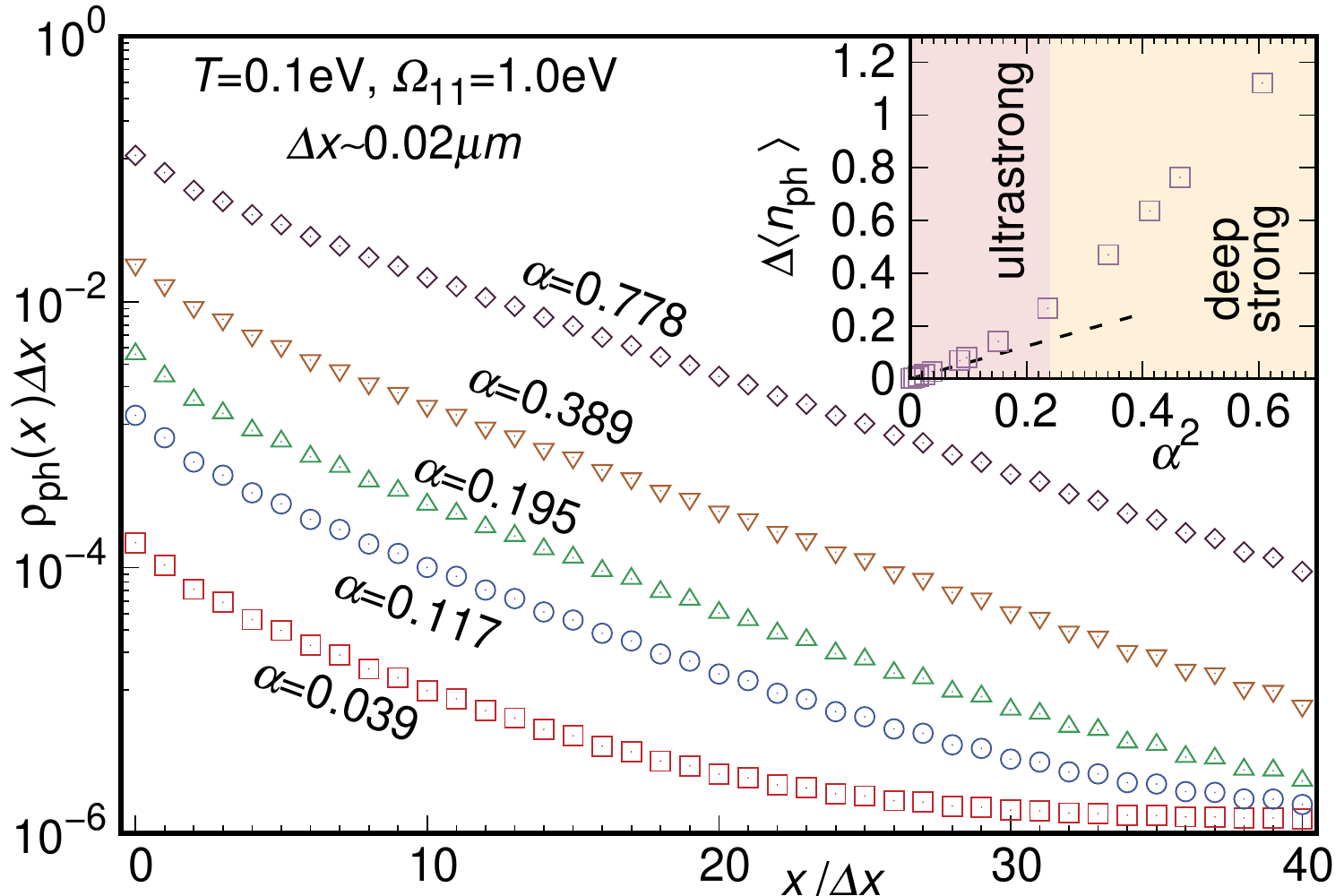}
	\caption{
Spatial distribution of the photon density $\rho_{\mathrm{ph}}(x)$ as a function of the distance $x$ to the dipole, for various coupling strengths $\alpha$, at $T=0.1$~eV and $\Omega_{11}=1.0$~eV. The inset shows the total bound photon number as function of the coupling $\alpha$.
	}
	\label{fig:photonDensity}
\end{figure}

The spin-photon coupling also renormalizes the photon propagator via the equation of motion,  $\mathcal{D}_{k\gamma,k'\gamma'}(iq_n) = \delta_{\gamma\gamma'}\delta_{kk'}\mathcal{D}^0_{k}(iq_n) - \frac{g_{k\gamma}g_{k'\gamma'}}{L}\mathcal{D}^0_{k}(iq_n)\chi_{\mathrm{sp}}(iq_n)\mathcal{D}^0_{k'}(iq_n)$, resulting in a photon bound state centered at the dipole. 
Figure~\ref{fig:photonDensity} shows the spatial distribution of the photon density $\rho_{\rm ph}(x)$ in the vicinity of the spin for various coupling strengths $\alpha$, and the total bound photon density $\Delta\langle n_{\rm ph} \rangle =\int dx [\rho_{\rm ph}(x)-\rho_0$], where $\rho_0$ is the noninteracting photon density due to thermal excitations. 
One can see that the photon distribution decays exponentially to the thermal background, and $\Delta\langle n_{\mathrm{ph}}\rangle$ becomes of order one in the deep strong coupling limit  ($\alpha\gtrsim 0.7$).

{\it Conclusions.}
We have introduced a 
vertex-based diagrammatic algorithm which allows to study strong light-matter coupling problems in the presence of a continuum of photon modes, and demonstrated its effectiveness with applications to the spin-boson model with coupling strength comparable to the cavity frequency and level splitting. 
The boldified~\cite{Prokofev2007,Prokofev2008,vanHoucke2012,Mishchenko2014,Deng2015,Rossi2018a,Rossi2018b,vanHoucke2019,vanHoucke2020} diagMC method~\cite{Prokofev1998,VanHoucke2010,Kozik:2010fla} has been reformulated to directly sample the four-point vertex with non-local-in-time interactions in pseudo-particle space~\cite{Barnes1976,Coleman1984}.
The self-consistency at the level of the triangular vertex was shown to improve the approximation at a given diagram order, and to speed up the convergence to the exact results.

With modified local propagators, the method can also be applied to the Anderson impurity model and related impurity problems with relevance for dynamical mean field theory. In this context, our method provides a systematic path for high-order, self-consistent {\it strong-coupling} expansions. While in equilibrium, alternative powerful Monte Carlo methods exist \cite{Werner2006,Weber2021}, the new approach introduced here is promising also for non-equilibrium applications, which will be the subject of forthcoming studies. 

{\it Acknowledgements.}
This work was supported by ERC Consolidator Grant No.~724103 (A.J.K., P.W.), by SNSF Grant No.~200021-196966, and the Marie Skłodowska Curie grant agreement No.~884104 (PSI-FELLOW-III-3i) (J.L.).  
K.L and M. E. were funded by the ERC Starting Grant No. 716648, and by the Deutsche Forschungsgemeinschaft (DFG, German Research Foundation) – Project-ID 429529648 – TRR 306 QuCoLiMa (``Quantum Cooperativity of Light and Matter'').  
The calculations have been performed on the Beo05 cluster at the University of Fribourg.

\bibliography{ref}

\onecolumngrid
\appendix

\newpage

\section{Quantization of the electromagnetic field inside the waveguide}
We considered a rectangular waveguide of width $a$ as shown in Fig.~1(a) of the main text.
The waveguide consists of perfectly conducting walls filled with a material of electric permittivity $\epsilon$ and magnetic permeability $\mu$.
In the absence of free charges and currents inside the waveguide, the homogeneous wave equation inside the waveguide is 
\begin{equation}
	\left(\nabla^2-\frac{1}{c^2}\partial^2_t\right)\bm{E}=0~,
	\label{eqn:waveEq}
\end{equation}
where $c$ is the speed of light.
Using the boundary conditions that enforce vanishing parallel components along the conducting surfaces, we can express the electric field 
\begin{equation}
	\bm{E}(\bm{r},t) = \sum^{}_{lm}\int dk_x\sum^{}_{j=1,2,3}\sqrt{\frac{\omega_k}{2\epsilon}}\bm{e}_j\left[ib^j_{lm}(k_x,t)v^j_{lm}(k_x,\bm{r})+c.c.\right]~,
	\label{eqn:Efield}
\end{equation}
as a linear combination of normalized mode functions
\begin{align}
	\stepcounter{equation}
	v^1_{lm}(k_x,\bm{r})&=\frac{2}{\sqrt{2\pi}a}e^{ik_xx}i\sin(k_ly)\sin(k_mz)~,\tag{\theequation a}\\
	v^2_{lm}(k_x,\bm{r})&=\frac{2}{\sqrt{2\pi}a}e^{ik_xx}\cos(k_ly)\sin(k_mz)~,\tag{\theequation b}\\
	v^3_{lm}(k_x,\bm{r})&=\frac{2}{\sqrt{2\pi}a}e^{ik_xx}\sin(k_ly)\cos(k_mz)~,\tag{\theequation c}
	\label{eqn:modeFunction}
\end{align}
with corresponding coefficients $b^j_{lm}(k_x,t)$~.
Due to the boundary conditions, the wavevector $k_l$ ($k_m$) along the $y$ ($z$) direction becomes the $l$th ($m$th) multiple of $\pi/a$ while $k_x$ takes continuous values.
The photon energy of the mode labeled by $l$ and $m$ with momentum $k_x$  along the $x$ direction equals $\omega_{k} = c\sqrt{k_x^2+k_l^2+k_m^2}$~.
Throughout the paper, we fix $\hbar=1$.

In order to quantize the transverse photon modes, we introduce the new momentum-dependent coordinate system
\begin{equation}
	\bar{\bm{e}}_{k\gamma} = \sum^{}_{j}O^{(k)}_{\gamma j}\bm{e}_{j}~,
	\label{eqn:coordinate}
\end{equation}
with the transformation matrix 
\begin{equation}
	O^{(k)} = \left(
		\begin{array}{ccc}
			\cos\theta_k\cos\phi_k & \cos\theta_k\sin\phi_k & -\sin\theta_k\\
			-\sin\phi_k & \cos\phi_k & 0\\
			\sin\theta_k\cos\phi_k & \sin\theta_k\sin\phi_k & \cos\theta_k
		\end{array}
	\right)~.
	\label{eqn:transform}
\end{equation}
The momentum index $k$ represents the three components $k_x$, $k_l$ and $k_m$, while the transformation parameters $\theta_k$ and $\phi_k$ are defined by
\begin{align}
	\stepcounter{equation}
	ck_x &= \omega_{lm}(k_x)\sin\theta_k\cos\phi_k\tag{\theequation a}~,\\
	ck_l &= \omega_{lm}(k_x)\sin\theta_k\sin\phi_k\tag{\theequation b}~,\\
	ck_m &= \omega_{lm}(k_x)\cos\theta_k\tag{\theequation c}~.
	\label{eqn:transformPara}
\end{align}

The corresponding new orthonormal mode functions are given by
\begin{equation}
	\bar{\bm{v}}_{k\gamma} = \sum^{}_{j}\bm{e}_jO^{(k)}_{\gamma j}v^j_k~,
	\label{eqn:neNormalMode}
\end{equation}
in which $\bar{\bm{v}}_{k3}$ is parallel to the wavevector $\bm{k}=(k_x,k_l,k_m)^{\intercal}$.
Note that $\bar{\bm{v}}_{k3}$ doesn't contribute to the mode expansion of $\bm{E}$ due to the divergence theorem $\nabla\cdot\bm{E}=0$~.

We now second-quantize the electric field by introducing two pairs of the photon creation and annihilation operators $a^{\dagger}_{k\gamma}$ and $a^{}_{k\gamma}$ for the two transverse modes $\gamma=1,2$:
\begin{equation}
	\bm{E}(\bm{r},t) = \sum^{}_{k\gamma}\sqrt{\frac{\omega_{k}}{2\epsilon}}\left[ia_{k\gamma}(t)\bar{\bm{v}}_{k\gamma}(\bm{r})+h.c.\right]~.
	\label{eqn:Efield}
\end{equation}
The resulting photon Hamiltonian reads 
\begin{align}
	\mathcal{H}_{\text{EM}}&=\frac{1}{2}\int_{}^{}d^3r\left[\epsilon\bm{E}^2+\frac{1}{\mu}\bm{B}^2\right] = \sum^{}_{k\gamma}\omega_{k}\left(a^{\dagger}_{k\gamma}a^{}_{k\gamma}+\frac{1}{2}\right)~.
	\label{eqn:Hphoton}
\end{align}

\section{Light-induced retarded spin-spin interaction}
Under the dipole gauge, the action of the spin-boson model can be written as 
\begin{equation}
	\mathcal{S}=\mathcal{S}_{\text{spin}}+\sum^{}_{k\gamma}\int_{0}^{\beta}d\tau~\bar{a}_{k\gamma}(\tau)\left[\partial_\tau+\omega_k\right]a_{k\gamma}(\tau) + \frac{1}{\sqrt{L}}\sum^{}_{k\gamma}\int_{0}^{\beta}d\tau~\left[\hat{\sigma}_1(\tau)g_{k\gamma}a_{k\gamma}(\tau)+c.c.\right]~,
	\label{eqn:effectiveAction}
\end{equation}
where $a_{k\gamma}(\tau)$ and $\bar{a}_{k\gamma}(\tau)$ are complex photon variables.
The coupling strength 
\begin{equation}
	g_{k\gamma}=ip\sqrt{\frac{\omega_{k}}{2\epsilon}}\bm{e}_1\cdot \bar{\bm{v}}_{k\gamma}~.
	\label{eqn:coupling}
\end{equation}
After integrating out the photon degrees of freedom, the effective spin action contains a retarded interaction between spins,
\begin{equation}
	\mathcal{S}_{\text{eff}}=\mathcal{S}_{\text{spin}}-\frac{1}{2}\int_{0}^{\beta}d\tau\int_{0}^{\beta}d\tau'~\hat{\sigma}_x(\tau)V(\tau-\tau')\hat{\sigma}_x(\tau')~.
	\label{eqn:effectiveAction}
\end{equation}
The form of the retarded spin interaction is
\begin{equation}
	V(\tau) = \frac{p_0^2}{\pi\epsilon a^2}\sum^{}_{\substack{lm\in \text{odd}\\\Omega_{11}<\omega_k}}\int_{-k_c}^{k_c}dk_x~\frac{\Omega_{11}^2}{\omega_k}\frac{\cosh\left(\omega_k\left(\tau-\frac{\beta}{2}\right)\right)}{\sinh\left(\frac{\beta\omega_k}{2}\right)}~,
	\label{eqn:retardedV}
\end{equation}
where $k_c=\frac{1}{c}\sqrt{\omega_c^2-\Omega_{11}^2}$, with the UV cutoff $\omega_c$ of the photon energy.

\section{Effective Rabi model}
\begin{figure}[t]
	\centering
	\includegraphics[width=1.0\textwidth]{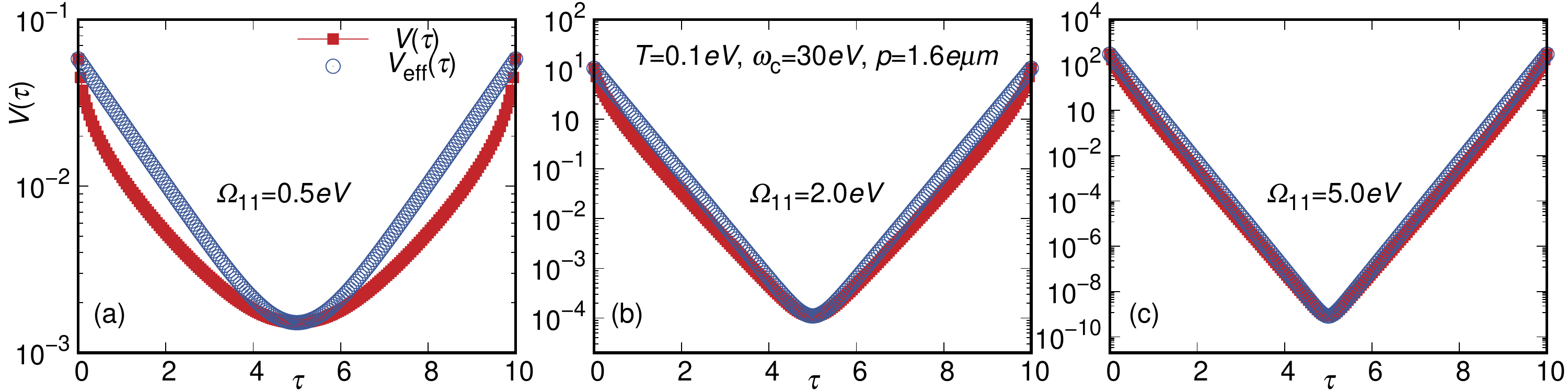}
	\caption{
		The retarded spin interaction $V_{\text{eff}}(\tau)$ of the effective cavity model fitted to the interaction of the waveguide model $V(\tau)$ for three photonic gap sizes of the waveguide: (a) $\Omega_{11}=$0.5~eV, (b) 2.0~eV, and (c) 5.0~eV.
		The temperature, size of the dipole moment, and UV cut-off frequency are fixed as 0.1~eV, 1.6~$e\mu m$, and 30~eV, respectively.
	}
	\label{fig:VtVsVeff}
\end{figure}

For the analysis and intuitive understanding of the waveguide spin-boson model, we introduce an effective Rabi model with a single photon mode
\begin{equation}
	\mathcal{H}_{\mathrm{eff}}=\frac{\Delta}{2}\hat{\sigma}_3 + g_{\mathrm{eff}}\hat{\sigma}_1\left(a^{\dagger}+a^{}\right) + \omega_{\mathrm{eff}}a^{\dagger}a^{}~,
	\label{eqn:HRabi}
\end{equation}
where $\hat{\sigma}_i$ ($i=1,2,3$) denotes the spin-$1/2$ Pauli operator, and $a^{\dagger}$ ($a^{}$) is the photon creation (annihilation) operator.
The effective parameters, the coupling strength $g_{\text{eff}}$ and the photon energy $\omega_{\text{eff}}$, are determined by fitting the retarded spin interaction of the waveguide,
\begin{equation}
	V(\tau) = 2\alpha^2\Omega_{11}^2\int_{\Omega_{11}}^{\omega_c}\frac{d\omega}{\sqrt{\omega^2-\Omega_{11}^2}}\frac{\cosh(\omega(\tau-\beta/2))}{\sinh(\omega\beta/2)}~,
	\label{eqn:Vwg}
\end{equation}
with the one of the Rabi model,
\begin{equation}
	V_{\text{eff}}(\tau) = g_{\text{eff}}^2\frac{\cosh\left(\omega_{\text{eff}}\left(\tau-\frac{\beta}{2}\right)\right)}{\sinh\left(\frac{\beta\omega_{\text{eff}}}{2}\right)}~.
	\label{eqn:Veff}
\end{equation}
Specifically, we identify $V_{\text{eff}}(\tau)$ with $V(\tau)$ at $\tau=0$ (or equivalently $\tau=\beta$) and $\tau=\beta/2$:
\begin{align}
	V(\tau=0) &= 2\alpha^2\Omega_{11}^2\int_{\Omega_{11}}^{\omega_c}d\omega~\frac{\coth\left(\frac{\beta\omega}{2}\right)}{\sqrt{\omega^2-\Omega_{11}^2}}=g_{\mathrm{eff}}^2\coth\left(\frac{\beta\omega_{\mathrm{eff}}}{2}\right)~,\label{eqn:Vt0fit}\\
	V(\tau=\beta/2) &= 2\alpha^2\Omega_{11}^2\int_{\Omega_{11}}^{\omega_c}d\omega~\frac{\mathrm{csch}\left(\frac{\beta\omega}{2}\right)}{\sqrt{\omega^2-\Omega_{11}^2}}=g_{\mathrm{eff}}^2\mathrm{csch}\left(\frac{\beta\omega_{\mathrm{eff}}}{2}\right)~. \label{eqn:Vb2fit}
\end{align}
Figure~\ref{fig:VtVsVeff} presents the resulting effective retarded interaction of the Rabi model (blue circles) and compares it to the one of the original waveguide model (red squares).
The functional form of the fitted interaction line gets closer to that of the waveguide model as we increase the photonic gap $\Omega_{11}$.
That implies that the effective Rabi model becomes a better approximation for a larger photonic gap.

For $T\ll\Omega_{11},\omega_{\mathrm{eff}}$ and $\omega_c\gg\Omega_{11}$, one can extract the asymptotic scaling of the effective parameters, $g_{\mathrm{eff}}$ and $\omega_{\mathrm{eff}}$.
In this limit, $\coth(\frac{\beta\omega}{2})\simeq \coth(\frac{\beta\omega_{\mathrm{eff}}}{2})\simeq 1$ and Eq.~(\ref{eqn:Vt0fit}) simplifies to 
\begin{equation}
	g_{\mathrm{eff}}^2 = 2\alpha^2\Omega_{11}^2\int_{\Omega_{11}}^{\omega_c}d\omega~\frac{1}{\sqrt{\omega^2-\Omega_{11}^2}}~.
\end{equation}
Using the integral
\begin{equation}
	\int_{\Omega_{11}}^{\omega_c}d\omega~\frac{1}{\sqrt{\omega^2-\Omega_{11}^2}}=\sinh^{-1}\left(\frac{\omega_c}{\Omega_{11}}\right)
	\label{eqn:integration1}
\end{equation}
and the asymptotic form of $\sinh^{-1}(\frac{\omega_c}{\Omega_{11}})\simeq\log\left(\frac{2\omega_c}{\Omega_{11}}\right)$ for $\omega_c\gg\Omega_{11}$, one finds
\begin{equation}
	g_{\mathrm{eff}}^2\simeq 2\alpha^2\Omega_{11}^2\log\left(\frac{2\omega_c}{\Omega_{11}}\right)~.
	\label{eqn:geff}
\end{equation}

On the other hand, $\mathrm{csch}(\beta\omega_{(\mathrm{eff})}/2)\sim 2e^{-\beta\omega_{(\mathrm{eff})}/2}$ in Eq.~(\ref{eqn:Vb2fit}) when $T\ll \omega_c$, so
\begin{equation}
	g_{\mathrm{eff}}^2e^{-\beta\omega_{\mathrm{eff}}/2} = 2\alpha^2\Omega_{11}^2\int_{\Omega_{11}}^{\omega_c}d\omega~\frac{e^{-\beta\omega/2}}{\sqrt{\omega^2-\Omega_{11}^2}}~.
\end{equation}
In the $\omega_c\gg\Omega_{11}$ limit,
\begin{equation}
	\int_{\Omega_{11}}^{\omega_c}d\omega~\frac{e^{-\beta\omega/2}}{\sqrt{\omega^2-\Omega_{11}^2}}\sim \int_{0}^{\infty}dx~\frac{e^{-\beta x/2}}{\sqrt{x(x+2\Omega_{11})}}=e^{\beta\Omega_{11}/2}K_0(\beta\Omega_{11}),
\end{equation}
where $K_n(z)$ is the modified Bessel function of the second kind.
In the $T\ll \Omega_{11}$ limit, $K_0(\beta\Omega_{11})\sim \sqrt{\frac{\pi}{2\beta\Omega_{11}}}e^{-\beta\Omega_{11}}$, which leads to the desired expression
\begin{equation}
	\omega_{\mathrm{eff}}\simeq \Omega_{11}\left[1+\frac{2}{\beta\Omega_{11}}\log\left\{\sqrt{\frac{2\beta\Omega_{11}}{\pi}}\log\left(\frac{2\omega_c}{\Omega_{11}}\right)\right\}\right]\sim\Omega_{11}~.
	\label{eqn:weff}
\end{equation}

\begin{figure*}[t]
	\centering
	\includegraphics[width=1.0\textwidth]{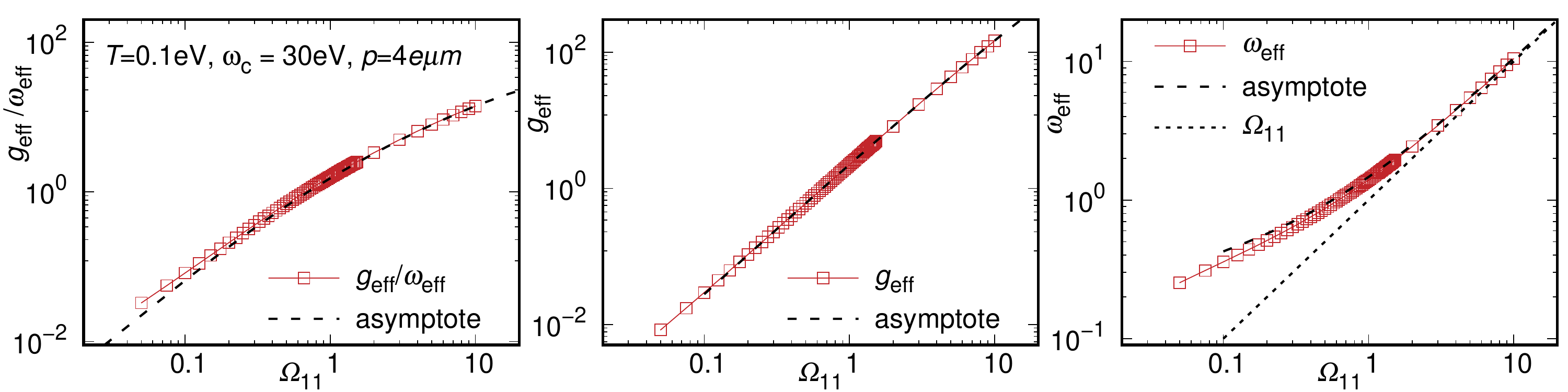}
	\caption{
		The parameters of the effective Rabi model as a function of the lowest photon energy $\Omega_{11}$ for the temperature $T=0.1$~eV and the dipole moment $p=4e \mu m$. The high-energy cutoff is fixed as $\omega_c=30$~eV.
	}
	\label{fig:effective_g_w}
\end{figure*}
Figure~\ref{fig:effective_g_w} shows the behavior of the effective parameters $g_{\mathrm{eff}}$, $\omega_{\mathrm{eff}}$, and their ratio $g_{\mathrm{eff}}/\omega_{\mathrm{eff}}$ as a function of the photonic gap $\Omega_{11}$ of the waveguide.
Those effective parameters approach the asymptotic formulae, Eq.~(\ref{eqn:geff}) and (\ref{eqn:weff}), namely $g_{\mathrm{eff}}\sim \Omega_{11}^2\log\left(2\omega_c/\Omega_{11}\right)$ and $\omega_{\mathrm{eff}}\sim \Omega_{11}$~.
We also observe that the ratio of the effective parameters $g_{\mathrm{eff}}/\omega_{\mathrm{eff}}$ monotonically increases as a function of $\Omega_{11}$ and asymptotically approaches $\sim\frac{p\Omega_{11}}{\sqrt{\varepsilon\pi^3c^3}}\log\left(\frac{2\omega_c}{\Omega_{11}}\right)$~.

By truncating the Hilbert space of the Rabi model to photon number $\le 1$, we can obtain an approximate analytic expression of the peak location of the lower polariton mode.
In the basis $\{|\!\uparrow;0\rangle,|\!\downarrow;0\rangle,|\!\uparrow;1\rangle,|\!\downarrow;1\rangle\}$ ($|\sigma_3;n\rangle$ denotes the spin $\sigma_3$ state with $n$ photons), the Hamiltonian of the Rabi model becomes
\begin{equation}
	\mathcal{H}_{\text{eff}}=\left(
		\begin{array}{cccc}
			\frac{\Delta}{2} & 0 & 0 & g_{\text{eff}}\\
			0 & -\frac{\Delta}{2} & g_{\text{eff}} & 0\\
			0 & g_{\text{eff}} & \omega_{\text{eff}}+\frac{\Delta}{2} & 0\\
			g_{\text{eff}} & 0 & 0 & \omega_{\text{eff}}-\frac{\Delta}{2}\\
		\end{array}
	\right)
	\label{eqn:Htoy}
\end{equation}
and the analytic expression of the eigenenergies is
\begin{equation}
	E = \frac{1}{2}\left\{\omega_{\text{eff}}\pm\sqrt{4g_{\text{eff}}^2+(\omega_{\text{eff}}\pm\Delta)^2}\right\}~.
	\label{eqn:eigenE}
\end{equation}
From Eq.~(\ref{eqn:eigenE}), we can track the peak location of the spin excitation spectrum by considering the Lehmann representation
\begin{equation}
	\chi_{\text{sp}}(\omega+i0^+)=\frac{1}{Z}\sum^{}_{mn}(e^{-\beta E_n}-e^{-\beta E_m})|\langle E_n|\sigma_1|E_m\rangle|^2/(\omega-E_n+E_m+i0^+)~.
	\label{eqn:Lehmann}
\end{equation}
In the limit $g_{\text{eff}},\omega_{\text{eff}}\gg \Delta$, the lowest energy peak location can be expressed as
\begin{align}
	\mathrm{min}_{m\neq n}|E_m-E_n| &= \sqrt{4g_{\text{eff}}^2+(\omega_{\text{eff}}+\Delta)^2} - \sqrt{4g_{\text{eff}}^2+(\omega_{\text{eff}}-\Delta)^2}~,\nonumber\\
	&= \frac{2\Delta}{\sqrt{1+\frac{\Delta^2+4g_{\text{eff}}^2}{\omega_{\text{eff}}^2}}} + \mathcal{O}(\Delta^2/g_{\text{eff}}^2,\Delta^2/\omega_{\text{eff}}^2)~,\nonumber\\
	&\sim \frac{2\Delta}{\sqrt{1+4g_{\text{eff}}/\omega_{\text{eff}}}}~.
	\label{eqn:asymptoticPeak}
\end{align}
This expression captures the red shift of the lower polariton mode as a function of the photonic gap.
As shown in Fig.~\ref{fig:effective_g_w}, the increase of the effective coupling strength $g_{\text{eff}}$ overcompensates the increase of $\omega_{\text{eff}}$ as a function of $\Omega_{11}$, leading to an overall increase of $g_{\text{eff}}/\omega_{\text{eff}}$~.
Furthermore, the lower polariton mode is inversely proportional to $g_{\text{eff}}/\omega_{\text{eff}}$~, as shown in the truncated subspace.
These results are consistent with the actual behavior of the lower polariton mode in the waveguide.

\section{Spin excitation spectrum for smaller photonic gap}
\begin{figure}[t]
	\centering
	\includegraphics[width=0.66\textwidth]{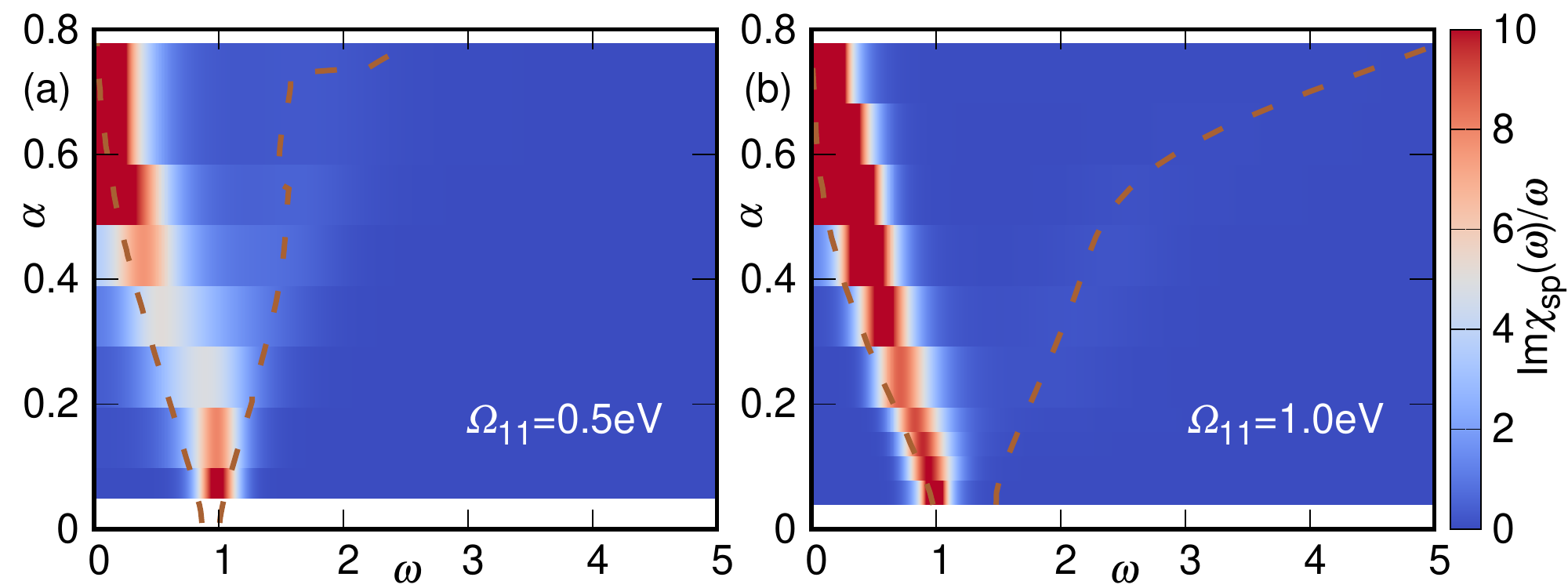}
	\caption{
		Spectral function of spin exciations for two different photonic gap size: (a) $\Omega_{11}=0.5$ and (b) $1.0$ where the bare spin excitation gap $\Delta=1.0$.
		$T=0.1$eV and $\omega_c=30eV$~.
		The (brown) dashed line shows the effective Rabi model prediction of the upper and lower polariton modes.
	}
	\label{fig:chiSpectrum}
\end{figure}
The size of the photonic gap $\Omega_{11}$ strongly modifies the spectral function of the upper polariton mode.
As we decrease $\Omega_{11}$, the peak location of the upper polariton mode is red shifted and the corresponding weight is significantly enhanced.
Figure~\ref{fig:chiSpectrum} presents the spin excitation spectrum $\mathrm{Im}\chi_{\mathrm{sp}}(\omega)/\omega$ for two different photonic gaps: $\Omega_{11}<\Delta$ and $\Omega_{11}=\Delta$, where $\Delta$ is the bare spin excitation energy.
In the small $\Omega_{11}$ case [Fig.~\ref{fig:chiSpectrum}(a)], significant spectral weight of the bare spin mode is transferred to the upper polariton mode in the ultra-strong coupling ($\alpha\gtrsim 0.3$).
However, for the larger photonic gap [Fig.~\ref{fig:chiSpectrum}(b)] the spectral weight of the bare excitation mostly resides in the lower polariton mode, while the upper polariton remains very weaks.
In both cases, the effective Rabi model captures the location of the upper and lower polariton modes.

\end{document}